\documentclass[trackchanges]{aastex7}

\usepackage{amsmath}

\begin{document}

\title{Upper limits on microhertz gravitational waves from supermassive black-hole binaries using PSR J1909\(-\)3744 data from the second IPTA data release}

\author[gname=Jing, sname=Zou]{Jing Zou}
\affiliation{Xinjiang Astronomical Observatories, Chinese Academy of Sciences, Urumqi 830011, China}
\affiliation{University of Chinese Academy of Sciences, Beijing 100049, China}
\affiliation{Institute of Optoelectronic Technology, Lishui University, Lishui 323000, China}
\email{zoujing@xao.ac.cn}  

\author[gname='Jingbo', sname=Wang]{Jingbo Wang}
\affiliation{Institute of Optoelectronic Technology, Lishui University, Lishui 323000, China}
\email[show]{1983wangjingbo@163.com} 

\author[gname='Jianping',sname=Yuan]{Jianping Yuan}
\affiliation{Xinjiang Astronomical Observatories, Chinese Academy of Sciences, Urumqi 830011, China}
\affiliation{Xinjiang Key Laboratory of Radio Astrophysics, Urumqi 830011, China}
\email[show]{yuanjp@xao.ac.cn}

\author[gname='De',sname='Zhao']{De Zhao}
\affiliation{Xinjiang Astronomical Observatories, Chinese Academy of Sciences, Urumqi 830011, China}
\email{zhaode@xao.ac.cn}

\author[gname='Yirong',sname=Wen]{Yirong Wen}
\affiliation{Xinjiang Astronomical Observatories, Chinese Academy of Sciences, Urumqi 830011, China}
\affiliation{University of Chinese Academy of Sciences, Beijing 100049, China}
\affiliation{Institute of Optoelectronic Technology, Lishui University, Lishui 323000, China}
\email{wenyirong@xao.ac.cn}

\author[gname=Wei, sname=Li, orcid=0009-0009-8247-3576]{Wei Li}
\affiliation{Xinjiang Astronomical Observatories, Chinese Academy of Sciences, Urumqi 830011, China}
\affiliation{University of Chinese Academy of Sciences, Beijing 100049, China}
\affiliation{Institute of Optoelectronic Technology, Lishui University, Lishui 323000, China}
\email{liwei@xao.ac.cn}  

\author[gname='Na',sname='Wang']{Na Wang}
\affiliation{Xinjiang Astronomical Observatories, Chinese Academy of Sciences, Urumqi 830011, China}
\affiliation{Xinjiang Key Laboratory of Radio Astrophysics, Urumqi 830011, China}
\email{ }

\author[gname='Yong',sname='Xia']{Yong Xia}
\affiliation{Xinjiang Astronomical Observatories, Chinese Academy of Sciences, Urumqi 830011, China}
\affiliation{University of Chinese Academy of Sciences, Beijing 100049, China}
\email{ }


\begin{abstract}

We present the results of a search for gravitational waves (GWs) from individual sources using high-cadence observations of PSR J1909\(-\)3744 obtained during an intensive observing campaign with the International Pulsar Timing Array second data release (IPTA-DR2) between July 2010 and November 2012. The observations, conducted at three different radio frequencies with the Nançay Radio Telescope (NRT) and Parkes Telescope (PKS) and five frequencies with the Green Bank Telescope (GBT), enabled precise corrections for dispersion measure effects and scattering variations. After these corrections, the timing residuals showed an unmodeled periodic noise component with an amplitude of 340 ns. Our analysis yields upper limits on the GW strain from individual sources, constraining it to be below \(1.9 \times 10^{-14}\) at 71 nHz and \(2.3 \times 10^{-13}\) at 1 \textmu Hz for average sky locations, while for optimal source locations the limits improve to \(6.2 \times 10^{-15}\) and \(8.9 \times 10^{-14}\) at the same frequencies, respectively. Our new limits are about a factor of 1.52 more stringent than those of Perera et al. based on an earlier EPTA data.

\end{abstract}

\section{INTRODUCTION}
Millisecond pulsars (MSPs) are old neutron stars that have been spun up to rotation periods of \(\leq\)30 milliseconds through mass accretion from binary companions—a process known as recycling \citep{Abdo_2013, Irr-2008-8}. These objects rank among the most stable rotators in the universe, with some exhibiting long-term rotational stability for over a decade, rivalling some of the most precise atomic clocks on Earth \citep[e.g.,][]{10.1093/mnras/stac3095, 10.1111/j.1365-2966.2011.20041.x}. This extraordinary rotational stability enables ultra-precise timing measurements, which are crucial for detecting gravitational waves (GWs) in the nanohertz regime. When organized into Pulsar Timing Arrays (PTAs), MSPs—with their sub-microsecond timing precision and multi-year observational baselines—form the most sensitive detectors of nanohertz-frequency gravitational waves \citep{Detweiler_1979,HD_1983}.

Supermassive black hole binaries (SMBHBs), thought to form via galaxy mergers, are expected to be abundant throughout galaxy evolution in the universe \citep{Burke_2019}. The superposition of GWs emitted by this cosmological population generates a stochastic gravitational wave background (SGWB), while particularly massive or nearby systems may stand out from this background as individually resolvable continuous wave sources \citep{Arzoumanian_2020, Sesana_2009}.  
For years, PTAs have placed increasingly stringent upper limits on the amplitude of the SGWB \citep[e.g.,][]{Arzoumanian_2016, Verbiest_2016, shannon_2015}. This effort culminated in a landmark achievement in June 2023, when four major collaborations—NANOGrav, EPTA, PPTA, and CPTA—jointly reported the first compelling evidence for a SGWB \citep{Agazie_2023, Antoniadis_2023, Reardon_2023, Xu_2023}. This result not only validates long-standing theoretical predictions but also opens the door to detecting individual supermassive black hole binaries through their continuous gravitational wave emissions.

PTAs enable long-term and high-cadence monitoring of exceptionally stable MSPs. Major PTAs include the European Pulsar Timing Array (EPTA), the Parkes Pulsar Timing Array (PPTA), the North American Nanohertz Observatory for Gravitational Waves (NANOGrav), the Chinese Pulsar Timing Array (CPTA), the Indian Pulsar Timing Array (InPTA), and the MeerKAT Pulsar Timing Array (MPTA) \citep{Antoniadis_2023, PPTA, NANOGrav, Xu_2023, InPTA, MeerPTA}. The International Pulsar Timing Array (IPTA)—a consortium that combines data from these regional PTAs—monitors an order of \(\sim\) 100 MSPs, with observational baselines ranging from \(\sim\)0.6 to 29.4 years. These datasets typically feature cadences of one to four weeks, occasionally supplemented by intensive observational campaigns \citep{IPTA_DR2}.

Building on previous efforts, substantial progress has been made in the search for both the stochastic gravitational wave background (SGWB) and continuous gravitational waves (CWs) from individual supermassive black hole binaries (SMBHBs). Several studies have placed upper limits on CW strain amplitudes using PTA data. For instance, based on the first data release from PPTA, \citet{Zhu_2014} constrained the CW strain amplitude to $\leq 1.7 \times 10^{-14}$ at 10 nHz. Using IPTA-DR2, \citet{Falxa_2023} reported a tighter upper limit of $\leq 9.1 \times 10^{-15}$ at the same frequency. Similarly, \citet{Arzoumanian_2023}, analyzing the NANOGrav 12.5-year dataset, derived a limit of $\leq 6.82 \times 10^{-15}$ at 7.65 nHz.

PTAs are sensitive to low-frequency (nanohertz) gravitational waves generated by SMBHBs with orbital periods of several years.
While most PTA searches have focused on the nanohertz regime, the microhertz (\textmu Hz) frequency range represents a largely unexplored window between the sensitivities of PTAs and space-based interferometers such as LISA. In this range, a variety of potential CW sources may exist. The most promising are SMBHBs nearing coalescence, whose orbital periods shorten from years to months, shifting their GW emission from the nanohertz to the microhertz domain \citep{Sesana_2013,Kelley_2018}. Intermediate-mass black hole binaries in dense stellar clusters may also radiate within this window \citep{Mandel_2018}, as well as compact stellar binaries with long orbital periods \citep{Korol_2017}. Moreover, exotic sources such as cosmic string oscillations \citep{Blanco-Pillado_2018} or boson cloud instabilities around rotating black holes \citep{Arvanitaki_2015} could produce quasi-monochromatic GW signals in this regime. Detecting or constraining such emissions would fill the frequency gap between PTA and LISA bands, offering new insight into the population and evolution of massive black hole binaries and testing gravitational physics across previously inaccessible timescales. Exploring this higher-frequency extension provides an opportunity to test PTA sensitivity limits and to constrain potential sources bridging the PTA-LISA gap.

High-cadence observations of individual pulsars have also yielded valuable upper limits. Using PSR B1937+21, \citet{Yi_2014} reported CW strain limits of $\leq 1.53 \times 10^{-11}$ and $\leq 4.99 \times 10^{-14}$ at a frequency of $10^{-7}$ Hz for random and optimal sky locations, respectively. Likewise, \citet{Perera_2018} found 95\% confidence upper limits of $\leq 3.5 \times 10^{-13}$ at 1 \textmu Hz and $\leq 1.4 \times 10^{-14}$ at 20 nHz for PSR J1713+0747. More recently, based on FAST observations of the same pulsar, \citet{Caballero_2025} derived a sky-averaged upper limit of $1.26 \times 10^{-12}$ at 1 \textmu Hz, and a directional limit of $4.77 \times 10^{-13}$.

Nevertheless, the accessible GW frequency range may be extended beyond conventional limits through advanced sampling strategies. For example, \citet{Wang_2021} proposed a staggered sampling approach to extend the frequency reach of PTAs beyond the nominal Nyquist limit. This method leverages the naturally asynchronous time-of-arrival (TOA) sampling of different pulsars to reconstruct GW signals at frequencies beyond what uniform sampling allows.

The GW detection window of PTAs is fundamentally determined by two observational parameters: the total observation span $T_\mathrm{obs}$ and the sampling cadence. The minimum detectable frequency is given by $f_\mathrm{min} \simeq 1/T_\mathrm{obs}$, as signal power below this scale is strongly correlated with intrinsic pulsar spin-down. The upper frequency limit is commonly assumed to be the Nyquist frequency, $f_\mathrm{Nyq} = N/(2T_\mathrm{obs})$, where $N$ is the total number of TOAs. However, for unevenly sampled data—a hallmark of PTA observations—this upper bound can be extended \citep{Yi_2014, Yardly_2010}.

For typical PTA data sets spanning decades with weekly to monthly cadence, the effective GW sensitivity band ranges from $\sim 10^{-9}$ Hz to $\sim 10^{-7}$ Hz \citep{PPTA}. The extended observational baselines enhance the PTA sensitivity at the lowest frequencies. While at high frequencies, PTA sensitivity is primarily limited by white noise, with the observing cadence setting a higher, less restrictive frequency ceiling. Although single-telescope observations rarely achieve sub-weekly cadence, coordinated multi-telescope campaigns can push the effective Nyquist limit into the microhertz regime through daily or sub-daily sampling.

Previous analyses of the complete IPTA-DR2 dataset have set stringent upper limits on the SGWB \citep{Antoniadis_2022} and searched for CWs \citep{Falxa_2023}. These studies maximize sensitivity at nanohertz frequencies. In contrast, our approach strategically sacrifices the long baseline for a high-cadence subset, enhancing sensitivity in the microhertz regime.
PSR J1909$-$3744 is among the most precisely timed MSPs. In IPTA-DR2, it has been observed by multiple telescopes from February 2004 to December 2014, with a total of 11,461 TOAs \citep{IPTA_DR2}. For this study, we selected a high-cadence subset covering July 2010 to November 2012, which contains 5,675 TOAs over a span of 873 days. This corresponds to an average interval of 3.1 days, with many sessions occurring daily or even multiple times per day. Such high-cadence sampling enhances the detectability of higher-frequency GWs, extending the effective search range into the microhertz regime.

High-cadence observations are essential not only for extending the detectable GW frequency range, but also for improving noise characterization. Since PTA data are inherently irregularly sampled, reducing cadence increases noise power across the entire frequency band. In this paper, we utilize high-cadence IPTA-DR2 data for PSR J1909$-$3744 to constrain the CW strain amplitude in the microhertz regime. PSR J1909$-$3744 is an ideal target for such a high-frequency search due to its exceptional long-term timing stability \citep{Liu_2020} and the dense observing cadence provided by the IPTA \citep{IPTA_DR2}.

The remainder of this paper is organized as follows: Section~\ref{sec:data} describes the observational data; Section~\ref{sec:timing} outlines the timing model and noise analysis; Section~\ref{sec:method} details the CW search methodology; Section~\ref{sec:result} presents the results; and Section~\ref{sec:summary} provides a summary and discussion.
\section{DATA} \label{sec:data}

\begin{figure*}[ht!]
\plotone{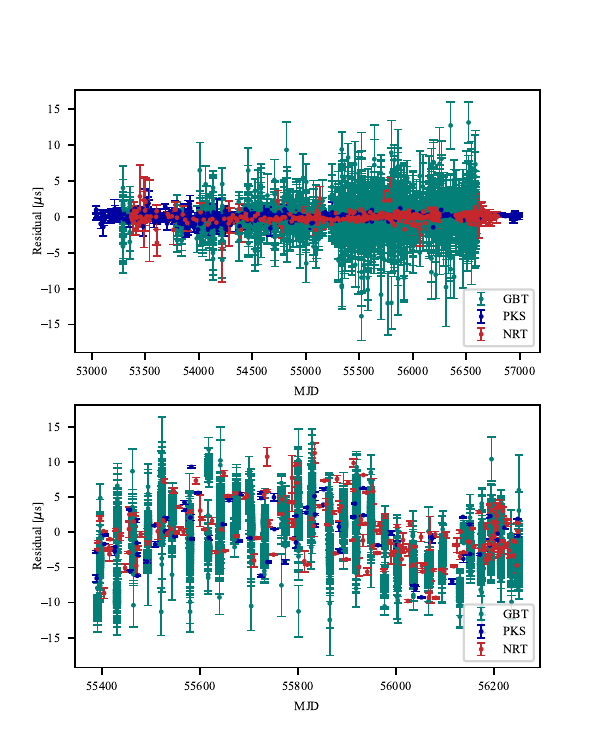}
\caption{The timing residuals of PSR J1909\(-\)3744. The weighted rms of the residuals is 1.20\(\ \mathrm{\mu} \)s. For clarity of the plot, the residuals of the different telescope are shown in separate panels. Data from the GBT, PKS, and NRT are shown in green, blue, and red, respectively. The top panel displays the residuals from the complete dataset, while the bottom panel shows the subset.} The average cadence of observations is 3.1 days across the data span between July 2010 and November 2012.
\label{fig:res}
\end{figure*}

The details of the observations are summarized in Table~\ref{tab:observation}. In our analysis, we utilized datasets from three telescopes, with their timing residuals illustrated in Figure~\ref{fig:res}.

The first dataset was obtained with the Nançay Radio Telescope (NRT) as part of the EPTA, using the BON (Berkeley-Orléans-Nançay) backend, which employs the ASP-GASP coherent dedispersion technique \citep{Desvignes_2016}. In August 2011, the L-band central frequency of the BON backend was shifted from 1400 MHz to 1600 MHz to accommodate the new wideband NUPPI backend \citep{Liu_2014}. The NRT dataset includes three observing frequencies: 1400, 1600, and 2000 MHz. Specifically, 156 TOAs were recorded at 1400 MHz from December 2004 to October 2011; 219 TOAs at 1600 MHz from September 2010 to December 2012; and 29 TOAs at 2000 MHz from March 2005 to August 2011.

\begin{deluxetable*}{lcccccc}
\tabletypesize{\normalsize}
\tablewidth{0pt} 
\tablecaption{Details of observations used in the analysis. The observation length varies and typical values for data sets are given. In total, 11461 observation epochs were include in complete data set and 5675 observation epochs were included in subset. The average cadence of observations is 3.1 days across the data span between July 2010 and November 2012. 
\label{tab:observation}}
\tablehead{
\colhead{Data set} & \colhead{Centre Freq (MHz)}& \colhead{Backend} & \multicolumn{2}{c}{Data span (MJD)} & \multicolumn{2}{c}{No. of TOAs} \\
\cline{4-5} \cline{6-7}
{ }&{ }&{ }& \colhead{complete data set} & \colhead{subset}  & \colhead{complete data set}& \colhead{subset}
} 
\startdata 
NRT &1400& BON & 53368\(-\)55843 & 55390\(-\)55843 & 156 & 26  \\
{}  &1600& BON & 55469\(-\)56794 & 55469\(-\)56264 & 219 & 123 \\
{}  &2000& BON & 53431\(-\)55796 & 55391\(-\)55796 & 29  & 12  \\
 \hline
PKS &700 &40CM\textunderscore PDFB3 &55055\(-\)55618&55404\(-\)55618&25 & 9   \\
{}  &{}  &50CM\textunderscore CPSR2 &54396\(-\)54989&\(-\)          &106& \(-\)   \\
{}  &1400&20CM\textunderscore PDFB1 &53548\(-\)54459&\(-\)          &77 & \(-\)   \\
{}  &{}  &20CM\textunderscore PDFB2 &54277\(-\)55276&\(-\)          &67 & \(-\)   \\
{}  &{}  &20CM\textunderscore PDFB3 &54754\(-\)55582&55388\(-\)55582&41 &11   \\
{}  &{}  &20CM\textunderscore PDFB4 &54752\(-\)55619&55385\(-\)55619&36 &9    \\
{}  &{}  &20CM\textunderscore CPSR2n&53431\(-\)54814&\(-\)          &63 &\(-\)    \\
{}  &{}  &20CM\textunderscore CPSR2m&53687\(-\)54224&\(-\)          &62 &\(-\)    \\
{}  &{}  &20CM\textunderscore WBCORR&53449\(-\)53530&\(-\)          &4  &\(-\)    \\
{}  &3100&10CM\textunderscore PDFB1 &53605\(-\)54465&\(-\)          &54 &\(-\)    \\
{}  &{}  &10CM\textunderscore PDFB2 &54294\(-\)55327&\(-\)          &41 &\(-\)    \\
{}  &{}  &10CM\textunderscore PDFB4 &54753\(-\)56992&55387\(-\)56249&173&63   \\
{}  &{}  &10CM\textunderscore WBCORR&53041\(-\)53819&\(-\)          &50 &\(-\)    \\
 \hline
GBT &800 & GUPPI&55278\(-\)56586 & 55396\(-\)56249 &4344 &2850 \\
{}  & {} & GASP& 53293\(-\)55278 & \(-\)           & 991 & \(-\)   \\
{}  &1200& GUPPI&55275\(-\)56598 & 55430\(-\)56250 & 787 & 472 \\
{}  &1400& GASP &53292\(-\)55492 & 55390\(-\)55492 & 698 & 21  \\
{}  &    & GUPPI&55275\(-\)56598 & 55430\(-\)56250 & 1246& 741 \\
{}  &1600& GUPPI&55275\(-\)56598 & 55430\(-\)56250 & 1057& 645 \\
{}  &1800& GUPPI&55275\(-\)56598 & 55430\(-\)56250 & 1135& 693 \\
\enddata
\tablecomments{In the table statistics, each frequency cut in NANOGrav is counted as a separate TOA, but when calculating the observation cadence, all frequency cuts from observations on the same day are treated as a single observation.}
\end{deluxetable*}

The second dataset was obtained with the Parkes radio telescope as part of the PPTA, using several backend systems including the Parkes Digital FilterBanks (PDFB1–PDFB4), the Caltech-Parkes-Swinburne Recorder version 2 (CPSR2), and the Pulsar Wideband Correlator (WBC). The PDFB systems are polyphase filter banks implemented with field-programmable gate arrays (FPGAs) and use 8-bit quantization. The WBC is a correlation spectrometer utilizing fast Fourier transforms (FFTs) on Canaris processors, with 2-bit (three-level) digitization. CPSR2 is a baseband recorder supporting coherent dedispersion over two 64-MHz dual-polarization channels; its successor APSR extends this capability to bandwidths up to 1000 MHz \citep{PPTA}. The Parkes dataset includes observations at 700, 1400, and 3100 MHz. At 700 MHz, 131 TOAs were recorded between October 2007 and February 2011 using the 40CM\textunderscore PDFB3 and 50CM\textunderscore CPSR2 backends. The 1400 MHz observations span March 2005 to February 2011, totaling 350 TOAs acquired using a combination of 20CM\textunderscore PDFB1–4, CPSR2m, CPSR2n, and WBCORR. At 3100 MHz, 318 TOAs were obtained from February 2004 to December 2014 using the 10CM\textunderscore PDFB1–2, PDFB4, and WBCORR systems.

The third dataset was collected with the Green Bank Telescope (GBT) under the NANOGrav project. Observations employed the Green Bank Ultimate Pulsar Processing Instrument (GUPPI) and the Green Bank Astronomical Signal Processor (GASP). GASP supports bandwidths up to 64 MHz, typically recorded in 60-second intervals over 4-MHz subbands. GUPPI offers broader bandwidth coverage (100–800 MHz), with data recorded in 1.5625-MHz subbands \citep{Arzoumanian_2015}. The GBT dataset covers five observing frequencies: 800, 1200, 1400, 1600, and 1800 MHz. The 800 MHz band spans October 2004 to October 2013 and includes 5335 TOAs using both GUPPI and GASP. For the period October 2004 to December 2013, 787 TOAs were recorded at 1200 MHz (GUPPI), 1944 at 1400 MHz (GUPPI and GASP), 1057 at 1600 MHz (GUPPI), and 1135 at 1800 MHz (GUPPI).

From the complete dataset, we extracted a high-cadence subset for focused analysis. For NRT, we selected observations from July 2010 to December 2012 at 1400, 1600, and 2000 MHz, totaling 161 TOAs. For Parkes, the subset covers the same period and at 700, 1400, and 3100 MHz, yielding 92 TOAs. For GBT, all five frequencies were included during this time span, resulting in 5422 TOAs. Our subsequent analysis focuses primarily on this high-cadence subset.

The complete dataset spans February 2004 to December 2014 and contains 11,461 TOAs in total. The high-cadence subset from July 2010 to November 2012 includes 5675 TOAs. Our timing solution is based on 278 unique observing days over an 873-day baseline, yielding an average inter-session interval of 3.1 days. For this calculation,  observations by different telescopes on the same day are counted as different session. Multiple TOAs obtained simultaneously in different frequency bands with the same telescope were counted as a single observing session. 

\section{TIMING THE PULSAR AND NOISE PROPERTIES} \label{sec:timing}

While a timing analysis of the complete dataset has already been published by \citet{IPTA_DR2}, our work focuses on a high-cadence subset. Initially, we used {\sc tempo2} \citep{Hobbs_2006} to generate a phase-coherent timing solution (see Table \ref{tab:timing}). However, because {\sc tempo2} utilizes a linearized model and a least-squares fitting algorithm, the presence of red noise can bias the resulting parameter estimates and their uncertainties \citep{Coles_2011,Caballero_2025}. To address this, we employed a Bayesian framework using the ENTERPRISE and ENTERPRISE\textunderscore EXTENSION software packages \citep{enterprise,enterprise_extension} to model the various noise components typically present in pulsar timing data. A comparison with the previous solution confirms that the timing parameters are consistent. Although the reduced precision resulting from the shorter data span is expected, it remains acceptable for GW analysis. Consequently, the timing solution derived from this 2-year high-cadence subset was used for the GW search.

\begin{deluxetable*}{lc}
\tabletypesize{\normalsize}
\tablewidth{0pt} 
\tablecaption{The timing model parameters of PSR J1909\(-\)3744 constrained using our data set given in Table \ref{tab:observation}. The binary parameters were measured using the T2 binary model given in TEMPO2. The position, spin frequency, and DM are given for the reference epoch of MJD 55820.
\label{tab:timing}}
\tablehead{
\multicolumn{1}{l}{Timing parameter} & \colhead{ }\\
} 
\startdata 
 Data span (MJD)& 55384-56257 \\
 Number of TOAs & 5675 \\
 Weighted rms timing residual (\(\mathrm{\mu} s\)) & 0.340\\
 & \\
 Right ascension (RA) (J2000) & 19\(\colon\)09\(\colon\)47.43178(9)\\ 
 Declination (DEC) (J2000) & \(-\)37\(\colon\)44\(\colon\)14.596(5)\\
 Proper motion in RA (mas/yr) & \(-\)9.4(2) \\
 Proper motion in DEC (mas/yr) & \(-\)35.7(4) \\
 Spin frequency, \(f\) \((\mathrm{s^{-1}}) \) & 339.3156871040759(2)\\
 Spin frequency \(1^\mathrm{st}\) time derivative, \(\dot{f}\)\((\mathrm{s^{-2}})\) & \(-\)1.614774(6)\(\times 10^{15}\)\\
 Reference epoch (MJD) & 55820\\
 Parallax, \(\pi\)(mas) & 0.89(2)\\
 Dispersion measure, DM \(\mathrm{cm^{-3} pc}\) & 10.39167(3)\\
 Dispersion measure \(1^\mathrm{st}\) time derivative, \(\dot{DM}\) \(\mathrm{(cm^{-3})}\) \(\mathrm{pc}\) \(\mathrm{yr^{-1}}\) & \(-\)0.0001(2)\\
 Dispersion measure \(2^\mathrm{nd}\) time derivative, \(\ddot{DM}\) \(\mathrm{(cm^{-3})}\) \(\mathrm{pc}\) \(\mathrm{yr^{-2}}\) & 3.6530(1)\\
 Orbital period, \(P_\mathrm{b}\) \((d)\) & 1.5334494753(1)\\
 Epoch of periastron, \(T_0\) (MJD) & 55015.4280907(1)\\
 Projected semi-major axis, \(x \) \((lt \text{-} s)\) & 1.8979909(6)\\
 Orbital inclination,  \(sin\ i\)  & 0.998(1)\\
 Companion mass, \(m_\mathrm{c}\) \( (M_{\odot})\) & 0.208(2)\\
  The electron density at 1 AU due to the solar wind, $n_{\rm e}$  & 4  \\
 & \\
 Clock correction procedure & TT (BIPM2015) \\
 Solar system ephemeris model & DE436 \\
 Units & TCB \\
\enddata

\end{deluxetable*}





In general, noise in a one-dimensional time series can be classified into two broad categories: time-uncorrelated (white) noise and time-correlated noise. The white noise model includes EFAC, a multiplicative scaling factor for TOA uncertainties; EQUAD, an additive term in quadrature accounting for pulse phase jitter; and ECORR, a term specifically modeling jitter noise correlated across subband TOAs within the observing bandwidth. As defined by \citet{Agazie_2023_2}, these three white noise terms come together with receiver/backend combination \(re/be\) dependence as
\begin{equation}
    C_{ij}=F^{2}(re/be)[\sigma^{2}_{S/N,i}+Q^{2}_{re/be}]\delta_{ij}+J^{2}(re/be)U_{ij}\ ,
    \label{eq:wn}
\end{equation}
where the \(\mathrm{i}, \mathrm{j}\) denote indices of TOA across all observing epochs, \(\delta_\mathrm{ij}\) is the Kronecker delta and F, Q, and J represent EAFC, EQUAD, and ECORR, respectively. ECORR is modeled using a blockdiagonal matrix, U, with values of 1 for TOAs from the same observation and 0 for all other entries.
We use a Uniform prior for EFAC and log-Uniform priors for both EQUAD and ECORR. Notably, the ECORR term is only included for data from the Green Bank Telescope (GBT), consistent with previous analyses by the NANOGrav collaboration \citep{NANOGrav}. Following the methodology of the IPTA-DR2 analysis \citep{IPTA_DR2}, we first estimate the white noise parameters and then hold their values fixed for the remainder of our analysis.

On the other hand, the time-correlated noise is primarily composed of two effects: achromatic red noise, a stochastic process typically attributed to pulsar spin irregularities, and chromatic noise due to dispersion measure (DM) variations as the signal propagates through the interstellar medium. To model the chromatic noise, we include a power-law DM variation model in addition to the standard linear (\(\dot{DM}\)) and quadratic (\(\ddot{DM}\)) DM terms in our timing model. Similarly, we use a power-law model to describe the achromatic timing noise. This power-law model is defined as:

\begin{equation}
    P = \frac{A^2}{12\pi^2}(\frac{f}{f_{yr}})^{-\gamma}  yr^3 ,
    \label{eq:rn}
\end{equation}
where A is the amplitude, \(\gamma\) is the spectral index and \(f_\mathrm{yr}\) is \(1/(1 \ \mathrm{year})\) in Hz.

Our analysis of PSR J1909\(-\)3744 indicates the presence of low-level achromatic red noise and stochastic DM noise in its timing data. Using the ENTERPRISE framework, we simultaneously model these noise components along with the pulsar's timing parameters. Both the red noise and the DM noise are modeled with power laws, using the prior settings detailed in Table \ref{tab:noise}.

\begin{deluxetable*}{cccc}
    \tabletypesize{\normalsize}
    \tablewidth{0pt} 
\tablecaption{Ranges and types for the priors used in the single-pulsar noise analysis and the results of PSR J1909\(-\)3744. The prior types Uni and log-Uni correspond to Uniform priors in linear and logarithmic space, respectively. With the median serving as the estimated value of the parameter result, set the error value at a 95\% confidence level.
\label{tab:noise}}
\tablehead{Parameter &Priors&Type& Value\\
} 
\startdata 
\(\mathrm{logA_{DM}} \) &\([-20,-11]\)& log-Uni &\( -13.80 \pm 0.73\) \\
\(\mathrm{\gamma_{DM}} \)&\([0,7]\)&Uni&\( 2.02 \pm 0.16\) \\
\(\mathrm{logA_{red}} \)&\([-20,-11]\)&log-Uni &\( -14.57 \pm 0.97\) \\
\(\mathrm{\gamma_{red}}\)&\([0,7]\)&Uni&\( 3.15 \pm 0.86\)\\
\enddata
\end{deluxetable*}

\begin{figure*}[ht!]
\plotone{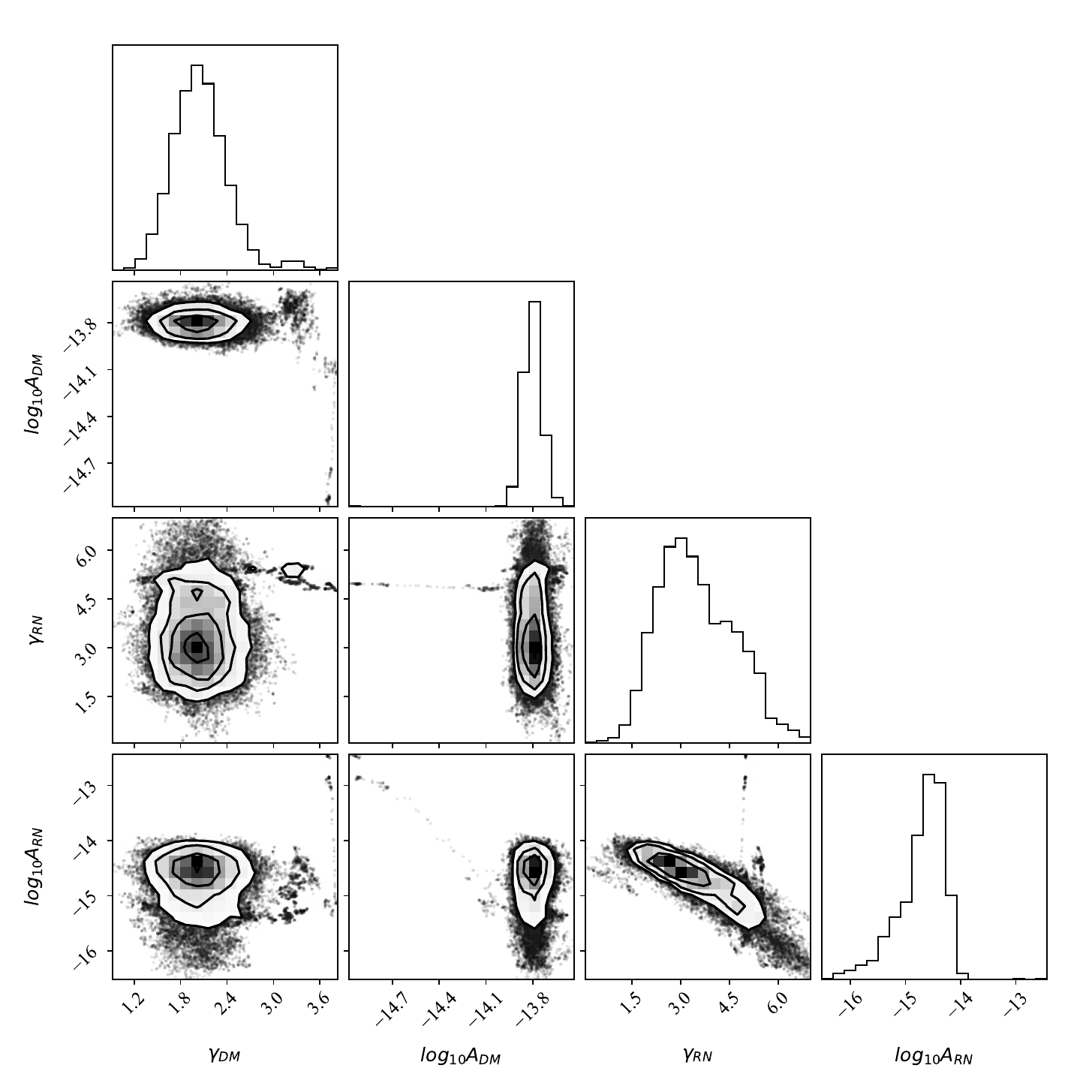}
\caption{2D marginalized posterior distribution for a subset of the noise parameters: DM stochastic noise component (logarithmic amplitude (\( \mathrm{log_{10}A}\)) and spectral index(\({\gamma}\))), and red noise component.}
\label{fig:noise}
\end{figure*}

The resulting noise parameters indicate that DM noise is characterized by an amplitude of \(\log_{10}A_{\mathrm{DM}} = -13.80 \pm 0.73\) and a spectral index of \(\gamma_{\mathrm{DM}} = 2.02 \pm 0.16\), while the red noise exhibits an amplitude of \(\log_{10}A_{\mathrm{red}} = -14.57 \pm 0.97\) and a spectral index of \(\gamma_{\mathrm{red}} = 3.15 \pm 0.86\) (see Table~\ref{tab:noise}).
When compared with the IPTA DR2 full data set\citep{IPTA_DR2} and NANOGrav\citep{Arzoumanian_2014} results, our measurements of red-noise and DM-noise amplitudes appear moderately smaller. This is consistent with expectations, as the shorter observational baseline in our data naturally limits sensitivity to the lowest-frequency noise components.
Bayesian model comparison yields a decisive Bayes factor of \(\log_{10}B = 5.95\), where \(B = \mathrm{evidence}[\mathrm{H_1}]/\mathrm{evidence}[\mathrm{H_0}]\) compares models with (\(\mathrm{H_1}\)) and without (\(\mathrm{H_0}\)) red and DM noise components. This result strongly supports the inclusion of both red noise and DM stochastic noise in our noise model.
Therefore,we adopted a noise model including both DM and red noise components in the subsequent analysis. A separate analysis without these components was not performed. To ensure our results is not biased by inadequate noise description, we did not perform a separate analysis without the DM and red noise term.
Figure~\ref{fig:noise} presents the corner plot showing the two-dimensional posterior correlations and one-dimensional marginalized distributions of the noise parameters. Both red noise and DM variations exhibit clear power-law spectral behavior.
In all subsequent analyses, our baseline noise model will include white noise (EFAC, EQUAD, and ECORR), red noise, and DM stochastic noise components.
For the complete data, while the same noise model components were considered, their specific parameter values were taken directly from \cite{IPTA_DR2}.

\section{METHODS} \label{sec:method}

GW signals from SMBHBs can be detected in pulsar timing residuals, provided that the pulsar has a sufficiently precise timing solution and the GW strain amplitude is strong enough to rise above the intrinsic noise. For circular binaries, the induced GW consists of two components: the high-frequency ``Earth term" and the low-frequency ``Pulsar term" \citep{Perera_2018,Jenet_2004}. 
The Earth term appears coherently across all pulsars, leading to correlated residuals that significantly enhance the signal-to-noise ratio (SNR), thereby improving detection sensitivity to weak GW signals. In contrast, the Pulsar term manifests independently in each pulsar due to the varying light travel times between the source and each pulsar. As a result, it captures the GW signal at an earlier epoch, encoding the historical evolution of the binary. This temporal diversity enables more accurate inference of the SMBHB parameters.In this work, we incorporate both the Earth and Pulsar terms in our analysis, with the goal of combining their advantages to achieve both high SNR and more precise parameter estimation.

We assess the sensitivity to individual GW sources using a Bayesian framework, following the methodology outlined in \citet{enterprise}. The noise properties of the pulsar are modeled using the \texttt{enterprise.signal} package, including the expected sinusoidal signature from a circular-orbit SMBHB. The model simultaneously fits for both deterministic and stochastic components: a sinusoidal GW signal, white noise terms (EFAC, EQUAD, and ECORR), and power-law models for dispersion measure (DM) variations and red noise. We marginalize over the pulsar timing model parameters and use the \texttt{PTMCMCSampler} to draw approximately \(1 \times 10^6\) posterior samples. For the power-law noise components, we adopt log-uniform priors on the amplitudes and uniform priors on the spectral indices. For the GW signal, we use uniform priors on the amplitude and phase, and a log-uniform prior on the GW frequency.

\subsection{signal model}

For a GW source located at right ascension \(\alpha\) and declination \(\delta\), we define the polar angle as \(\theta = \pi/2 - \delta\) and the azimuthal angle as \(\phi = \alpha\). The strain induced by such a source can be expressed in terms of two polarization components as follows:

\begin{equation}
    h_{ab}(t,\hat\Omega) = e^+_{ab}(\hat \Omega)h_+(t,\hat \Omega) + e^\times _{ab}(\hat \Omega)h_\times(t, \hat \Omega)
    \label{eq:h_ab}
\end{equation}
where \(\hat{\Omega}\) denotes the unit vector pointing from the GW source to the Earth (i.e., the direction of wave propagation), \(h_{+,\times}\) represents the two polarization amplitudes, and \(e^{+}_{ab}\) and \(e^{\times}_{ab}\) are the corresponding polarization tensors. In the Solar System Barycenter (SSB) frame, these tensors are given by:
\begin{equation}
    \begin{split}
    e^+_{ab}=\hat{p_a}\hat{p_b} - \hat{q_a}\hat{q_b} \ \\
    e^{\times}_{ab}=\hat{p_a}\hat{p_b} + \hat{q_a}\hat{q_b},
    \label{eq:e_ab}
    \end{split}
\end{equation}
and are constructed from basis vectors
\begin{equation}
    \begin{split}
        \hat{n} &=(sin\theta cos\phi,sin\theta sin\phi, cos\theta)= -\hat{\Omega} \\
        \hat{p} &=(cos\psi cos\theta cos\phi - sin\psi sin\phi, cos\psi cos\theta sin\phi + sin\psi cos\phi, -cos\psi sin\theta) \\
        \hat{q} &=(sin\psi cos\theta cos\phi + cos\psi sin\phi, sin \psi cos\theta sin\phi- cos\psi cos\phi, -sin\psi sin\theta),
    \end{split}
\end{equation}
where \(\psi\) is the GW polarization angle. It is worth noting that adopting this basis ensures consistency with \citet{Arzoumanian_2023} and aligns with conventions used by other GW detectors. The antenna pattern functions \(F^{+,\times}(\hat{\Omega})\), which describe the response of a pulsar (with unit position vector \(\hat{u}\)) to gravitational wave emission, are constructed from these polarization tensors. Following the formulation in \citet{Taylor_2016}, they are given by:
\begin{equation}
    F^A(\hat{\Omega}) \equiv \frac{1}{2}\frac{\hat u^a \hat u^b}{1+\hat{\Omega}\cdot \hat u} e^A_{ab}(\hat{\Omega}).
    \label{eq:F}
\end{equation}

The signal \(s\) induced by the GW, as seen in the pulsar’s residuals, can now be written as:
\begin{equation}
    s(t,\hat \Omega)=F^+(\hat \Omega) \Delta s_+(t) + F^\times (\hat \Omega) \Delta s_\times (t),
    \label{eq:s}
\end{equation}
where \(\Delta s_{+,\times}\), representing the difference between the signal induced at the Earth (the Earth term) and at the pulsar (the Pulsar term), can be expressed as:
\begin{equation}
    \Delta s_{+,\times}(t)=s_{+,\times}(t_p) - s_{+, \times}(t),
    \label{eq:delta_s}
\end{equation}
where \(t\) and \(t_\mathrm{p}\) denote the times of GW passage at the Earth and the pulsar, respectively. These times are geometrically linked by:
\begin{equation}
    t_p = t - L(1 + \hat{\Omega} \cdot \hat u),
    \label{eq:t_p}
\end{equation}
where \(\hat{u}\) indicates the line of sight vector to the pulsar, while \(L\) signifies the distance to the pulsar.

We can write \(s_{+, \times}\) for a circular binary at zeroth post-Newtonian (0-PN) order as:
\begin{equation}
    \begin{split}
        s_+(t) &= \frac{M^{5/3}}{d_L\omega(t)^{1/3}}[-sin 2\Phi (t)(1 + cos^2 \iota)], \\
        s_\times (t) &= \frac{M^{5/3}}{d_L\omega(t)^{1/3}}[2cos 2\Phi (t)cos \iota],
    \end{split}
    \label{eq:s2}
\end{equation}
where \(\iota\) corresponds to the inclination angle of the SMBHB, \(d_L\) represents the luminosity distance to the source, \(\omega(t)\) and \(\Phi (t)\) are the time-dependent angular orbital frequency and phase, respectively, and \(M \equiv (m_1 m_2)^{3/5} / (m_1 + m_2)^{1/5}\)  is known as the chirp mass, representing a combination of the two black hole masses. The variables \(M\) and \(\omega\) refer to the redshifted values of these quantities, relating to the rest-frame versions \(M_\mathrm{r}\) and \(\omega_\mathrm{r}\) as 
\begin{equation}
    \begin{split}
        M_r &=\frac{M}{1 + z},\\
        \omega_r &= \omega(1+z).
    \end{split}
    \label{eq:M}
\end{equation}
However, PTAs are currently considered sensitive only to individual SMBHBs limited to the local universe, in the vicinity where \((1+z) \sim 1\).

For a CW, the initial orbital angular frequency \(\omega_0\), which corresponds to \(\omega_0 = \omega(t_0)\), is related to the GW frequency by \(\omega_\mathrm{0} = \pi f_\mathrm{GW}\). Defining the reference time \(t_0\) as MJD 0 for this search, the time-dependent orbital phase and frequency of the binary are given by
\begin{equation}
    \begin{split}
        \Phi(t) &= \Phi_0 + \frac{1}{32}M^{-5/3}[\omega_0^{15/3}- \omega(t)^{-5/3}], \\
        \omega(t)&=\omega_0(1-\frac{256}{5}M^{5/3}w_0^{8/3}t)^{-3/8},
    \end{split}
    \label{eq:phi}
\end{equation}
where \(\Phi_0\) represents the initial orbital phase.

\subsection{Bayesian Method}

\begin{deluxetable*}{ccc}
    \tabletypesize{\normalsize}
    \tablewidth{0pt} 
\tablecaption{
The prior ranges and types employed in the Bayesian single source GW upper limit analysis. The prior types “Uni” and “log-Uni” denote uniform priors in linear and logarithmic spaces, respectively. An asterisk (*) following the celestial coordinates \(\phi\) and cos\(\theta\) indicates the total coverage range listed, but each analysis utilizes only the coordinate ranges corresponding to each sky cell. The \(f_\mathrm{GW}\) is fixed separately for each individual analysis, with its value restricted within the prior range specified.
\label{tab:gw}}
\tablehead{Parameter & Priors & Type\\
} 
\startdata 
\(\mathrm{A_\mathrm{GW}} \) & \([10^{-18},10^{-11}] \) & Uni\\
\( \phi \)&\([0,2\pi]\) & Uni \\
\( \mathrm{cos\theta}\)&\( [-1,1]\)& Uni \\
\(\mathrm{f_\mathrm{GW}}\)&\( [5.0 \times 10^{-9}, 3.2 \times 10^{-6}]\)& Uni/fixed from grid \\
\(\mathrm{\iota}\) & \([0,2\pi]\) & Uni \\
\(\mathrm{\psi}\) & \([0,\pi]\) & Uni \\
\(\mathrm{\phi_0}\) & \([0,2\pi]\)& Uni\\
\enddata
\end{deluxetable*}

The posterior distributions of GW parameters were determined through Bayesian inference. Consistent with previous CW searches \citep{Falxa_2023,Arzoumanian_2023}, we used a Markov Chain Monte Carlo (MCMC) sampler\citep{PTMCMC}. This procedure was preceded by a prior-recovery analysis, performed to guarantee that the sampler could search the entire prior volume prior to our data analysis. 

The CW signal model can be described by nine global parameters:
\begin{equation}
    \{\theta,\phi,f_\mathrm{GW},\Phi_0,\psi,\iota,M,d_L,h_0\},
    \label{eq:param}
\end{equation}
which describe the circular SMBHB's: position on the sky(\(\theta,\phi\)); GW frequency, related to the orbital frequency at a reference time (\(f_\mathrm{GW}\)); orbital phase at a reference time (\(\Phi_0\)); GW polarization angle (\(\psi\)); orbital inclination (\(\iota\)); chirp mass (\(M\)); luminosity distance (\(d_\mathrm{L}\)); strain amplitude (\(h_0\)), which is related to the chirp mass, GW frequency, and luminosity distance.

Since \(h_0\) can be defined as
\begin{equation}
    h_0 = \frac{2M^{5/3}(\pi f_\mathrm{GW})^{2/3}}{d_L},
    \label{eq:h0}
\end{equation}
there is a degeneracy between \(h_0\), \(M\), \(f_\mathrm{GW}\), and \(d_\mathrm{L}\), and therefore only eight of these parameters are required to fully describe the global GW signal. The specific prior distributions adopted for these GW signal parameters are detailed in Table \ref{tab:gw}. The analysis frequency band, which spans from 3.2 nHz to 3.2 \textmu Hz, was divided into 61 equally spaced bins. Within each bin, we derived 95\% upper limits on the strain amplitude for potential GW sources by taking the 95th percentile of the relevant strain posterior distributions. The error associated with these limits is calculated as follows:
\begin{equation}
    \sigma_{UL} = \frac{\sqrt{x(1-x)/N_s}}{p(h_0=h^{95 \%}_0 | D)},
    \label{eq:sigma}
\end{equation}
where \(x=0.95\) and \(N_\mathrm{s}\) is the number of effective samples in the MCMC chain.

\subsection{All-sky Upper Limits and Sky Map}

When detecting GWs potentially emanating from SMBHBs at any sky location, we adopt Uniform priors for the source sky coordinates (\(cos \theta , \phi\)), the cosine of the source inclination  cos\(\iota\), the polarization angle \(\psi\), and GW phase \(\Phi_0\). To guarantee the derivation of the most conservative upper limits, Uniform priors on \(h_0\) are applied specifically for this purpose. In the course of this analysis, the prior distribution for \(\mathrm{log_{10}(h_0)}\) covers the interval \([-18,-11]\). This interval not only accounts for a conservative region surrounding the sensitivity level of the latest data sets (of order \(-\)15) but also has a lower bound significantly beneath our instrumental sensitivity.

Our searches cover the microhertz GW band, with frequency limits set by the data's temporal properties. The lower bound is determined by the total observation span (\(f_{min} = 1/T_{obs}\)), while the upper bound is set by the observation cadence. For the complete dataset, which spans 10.83 years, the theoretical minimum frequency is \(f_\mathrm{GW} = 3.0 \times 10^{-9} \ \mathrm{Hz}\). The maximum frequency, dictated by a 3.1-day cadence, is \(f_\mathrm{GW} = 3.7 \times 10^{-6} \ \mathrm{Hz}\).
We performed searches at multiple frequencies within this range. To better visualize the sensitivity trend at the lowest frequencies, we used \(f_\mathrm{GW} = 3.2 \times 10^{-9} \ \mathrm{Hz}\) as the starting point for our sensitivity curve characterization.


The sensitivity of an individual pulsar differs in different directions in the sky. Following detection analysis, we assessed sensitivity variations by placing upper limits on 768 isotropically distributed sky pixels (53.72 square degrees each), calculated using healpy \citep{Zonca2019}. This grid resolution balances the need for detailed sky sensitivity mapping with computational feasibility and the expected localization performance of J1909\(-\)3744. A Uniform prior across all 768 pixels was assigned to the sampler to facilitate full sky exploration within the analysis.

Due to the substantial computational cost associated with performing 768 independent runs, the sky map was generated at a single frequency, and only upper limits were computed. The frequency of \(7.1 \times 10^{-8}\ \mathrm{Hz}\) was selected for the sky map because it yielded the highest sensitivity in the sky-averaged analysis. Meanwhile, the frequency of  \(1.0 \times 10^{-6} \ \mathrm{Hz}\) was chosen for the upper limit calculation as it is of particular physical interest in this research as well as to facilitate a direct comparison with previous work.

\section{RESULT} \label{sec:result}

\subsection{All-sky Upper Limits}

\begin{figure*}[ht!]
\plotone{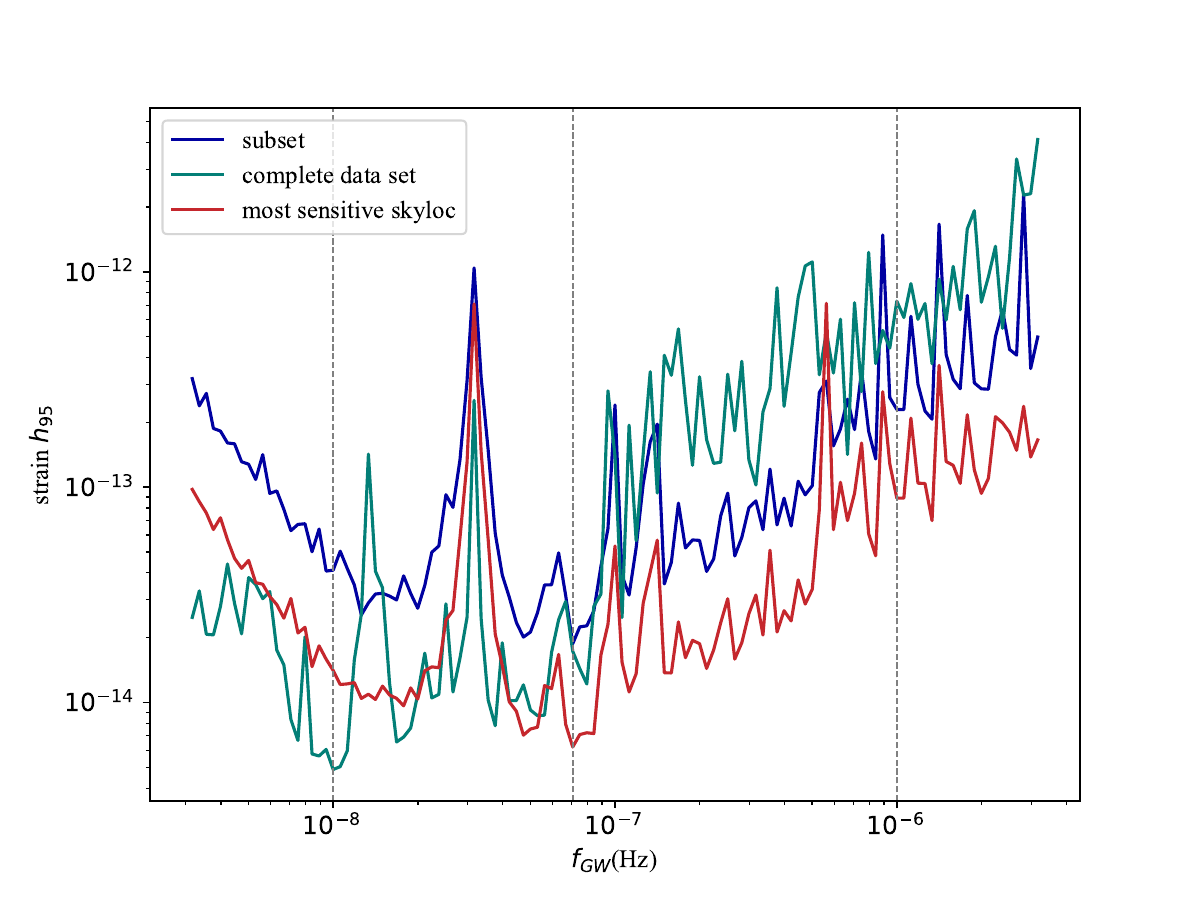}
\caption{
The 95 \(\%\) upper limits on the GW strain amplitude produced by SMBHBs based on the timing observations of PSR J1909\(-\)3744. The blue, and green lines represent GW sky-averaged sensitivity. The blue curve and red curve are derived from a partial data set, whereas the green curve reflects the complete data set. The red line shows sensitivity for the most sensitive sky location.} 
\label{fig:result}
\end{figure*}

The single pulsar GW search analysis cannot conclusively detect GWs, but instead produces upper limits on strain amplitudes \citep{Perera_2018}. Thus, we proceeded to place all-sky upper limits on the GW strain, with results shown in Figure \ref{fig:result}. The analysis was performed using the specified model (WN+RN+DM+CW).
The blue curve represents sky-averaged obtained from the subset data, achieving its optimal sensitivity of \(1.9\times 10^{-14}\) at 71 nHz. However, the Following the convention used in previous studies, we also quote the sensitivity at upper limit  at 1 \textmu Hz as a representative value in the microhertz regime;at this frequency, the sky-averaged upper limit is \(2.3 \times 10^{-13}\). The red curve presents the results for the the most sensitive sky location, which will be discussed in Section \ref{sec:sky-map}.

To verify the rationality of our results, we perform comparative analysis using the complete observational data set (green curve) in Figure \ref{fig:result}). This dataset achieved its optimal sensitivity of \(4.9\times 10^{-15}\) at 10 nHz. Note that the green curve exhibits a narrow peak near $f \sim 10^{-8}\,\mathrm{Hz}$. This feature is not indicative of a CW signal. A dedicated Bayesian model comparison at this frequency yields a Bayes factor of $log_{10}B_{\rm CW/N} = 1.7$, which provides no statistical support for the presence of a signal. Similar frequency-localized spikes have been reported in previous PTA CW searches, including analyses based on IPTA DR2 \citep{Falxa_2023} and other data sets. At our target frequency of 1 \textmu Hz, however, the sensitivity was reduced to \(7.3 \times 10^{-13}\). The performance shows a distinct frequency dependence: the complete dataset demonstrates superior sensitivity in the lower frequency regime (below \(\sim\) 100 nHz), while the timing subset shows enhanced performance at higher frequencies (above \(\sim\) 0.5 \textmu Hz). This dichotomy stems from fundamental differences in the datasets observational parameters. The timing subset, characterized by its higher cadence, provides more frequent sampling of the TOAs, thereby increasing its sensitivity to the more rapid variations induced by higher-frequency GWs. Conversely, the complete dataset benefits from its extended temporal baseline (longer observation time), which enables better characterization and integration of the slower, cumulative effects produced by lower-frequency GW signals. This complementary behavior precisely aligns with theoretical expectations for pulsar timing array sensitivity, where low-frequency detection capability scales with observation duration while high-frequency response improves with sampling rate. The results quantitatively demonstrate how strategic dataset selection can be optimized for specific GW frequency ranges of interest.

\subsection{Sky Map} \label{sec:sky-map}
\begin{figure*}[ht!]
\plotone{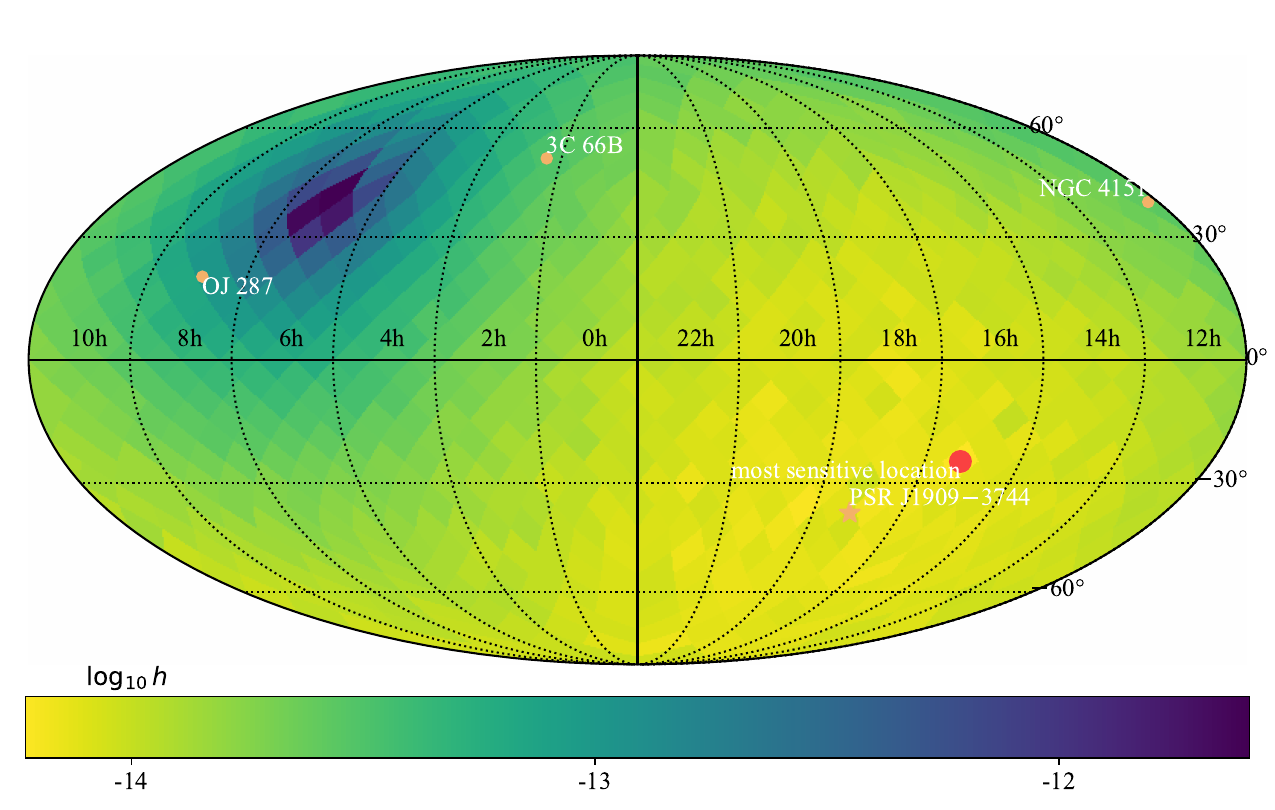}
\caption{
Sky map of  95\% upper limits on the GW strain amplitude produced
by SMBHBs based on the timing observations of PSR J1909\(-\)3744 at \(f_\mathrm{GW} = 71  \mathrm{nHz}\) GW freqency. Pulsar location is shown as yellow stars. yellow diamonds indicate the positions of three known SMBHB candidates that could contain an SMBHB. The most sensitive pixel is marked with a red dot, and is located at an RA of \(\mathrm{17^h 14^m 58^s}\) and a DEC of \(-27^{\circ} 19^{'} 35^{''}\). While our single-pulsar analysis achieves optimal sensitivity near the pulsar position, it shows negligible response to the antipodal sky location (anti-pulsar direction).}
\label{fig:70nHz}
\end{figure*}

Figure \ref{fig:70nHz} shows how the GW strain upper limits vary across the sky for a subset of sources, evaluated at the most sensitive CW frequency of \(f_\mathrm{GW} = 7.1 \times 10^{-8} \ \mathrm{Hz}\). The map reveals that, as expected, the least sensitive region is located in the anti-direction of the pulsar.

At the most sensitive location, the strain upper limit is \(h_0 < 5.9 \times 10^{-15} \), while at the least sensitive location, \(h_0 < 2.6 \times 10^{-12} \), indicating a sensitivity variation by a factor of \(\sim 435\) across the sky. We found that the most sensitive region is located near the pulsar at RA of \(\mathrm{17^h 14^m 58^s}\) and DEC of \(\mathrm{-27^{\circ} 19^{'} 35^{''}}\). We adopt this location to generate the sensitivity curve (red line) presented in Figure \ref{fig:result}. Clearly, the sensitivity curve obtained at the most sensitive location significantly outperforms that derived from the average sky location. At the most sensitive frequency of 71 nHz, we constrain the 95\% confidence interval for the strain amplitude to \(6.2 \times 10^{-15}\), while at 1 \textmu Hz the result is \(8.9 \times 10^{-14}\).

\begin{figure*}[ht!]
\plotone{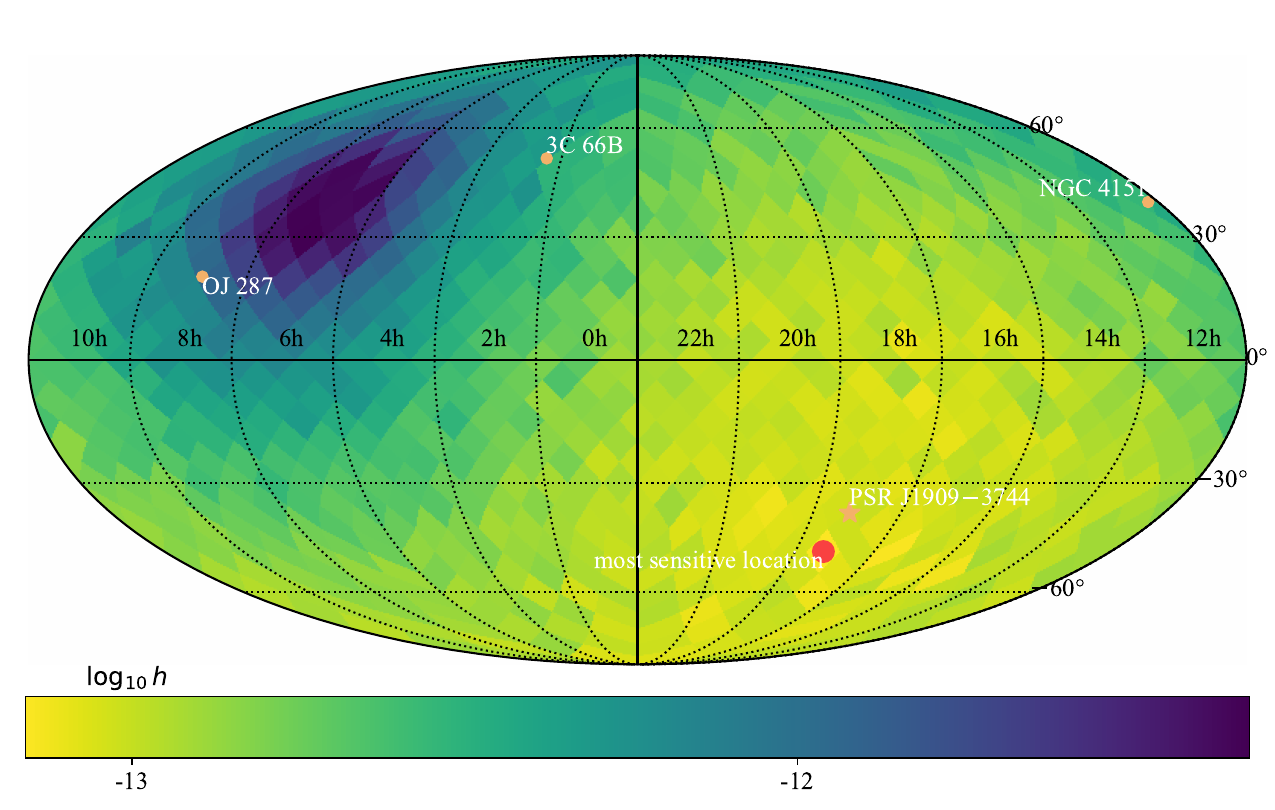}
\caption{
Sky map of  95\% upper limits on the GW strain amplitude produced
by SMBHBs based on the timing observations of PSR J1909\(-\)3744 at \(f_\mathrm{GW} = 1 \mu\mathrm{Hz}\) GW freqency. The most sensitive pixel is marked with a red dot, and is located at an RA of \(\mathrm{19^h 17^m 8^s}\) and a DEC of \(-32^{\circ} 19^{'} 00^{''}\).}
\label{fig:1mu}
\end{figure*}

In Figure \ref{fig:1mu}, the 1 \textmu Hz analysis reveals the characteristic dipolar sensitivity pattern expected for single-pulsar GW detection, with strain upper limits ranging from \(h_0 < 6.9 \times 10^{-14} \) at the most sensitive pixel to \(h_0 < 4.8 \times 10^{-12}\) at the least sensitive pixel (a variation factor of \(\sim 78\)). This confirms that the geometric sensitivity effect persists across different frequency bands. Furthermore, the upper limit at the most sensitive point is approximately one order of magnitude lower than that found at 71 nHz.

\section{Discussion and Summary}  \label{sec:summary}

We used high-cadence observations of PSR J1909\(-\)3744 to place upper limits on the strain amplitude of CWs produced by individual SMBHBs in circular orbits.
Based on Bayesian method, we found a 95\% upper limit on the sky-averaged strain amplitude of CWs to be \(2.3 \times 10^{-13}\) at a reference frequency of 1 \textmu Hz. For an optimal sky location, the 95\% upper limit on the sky-averaged strain amplitude of CWs to be \(8.9 \times 10^{-14}\) at the same reference frequency. We also found that our timing results place upper limits on the sky-averaged and optimal strain amplitude of low frequency CWs to be \(1.9 \times 10^{-14}\) and \(6.2 \times 10^{-15}\) at a reference frequency of 71 nHz, respectively. Compared to the upper limits on GWs from PSR J1713+0747 reported by \citet{Perera_2018} and \citet{Caballero_2025}, our study demonstrates that PSR J1909\(-\)3744 improves the sky-averaged and optimal-location upper limits at 1 \textmu Hz by factors of 1.53 and 5.49, respectively. 
In the microhertz band, the sensitivity is primarily determined by the high-frequency noise characteristics and the temporal sampling of individual pulsars. This regime corresponds to systems at significantly later stages of the SMBHB inspiral than the nanohertz band probed by conventional PTA analyses.

Our results were enabled by the use of high-cadence observations from the IPTA, which combines data from the  PPTA, EPTA and NANOGrav. The combination of datasets from multiple PTAs provides a significantly shorter observational cadence than any individual PTA alone. This enhanced cadence improves our constraints on the continuous GW strain amplitude at microhertz frequencies which were more stringent than those obtained in previous single-PTA studies.

However, a single pulsar cannot be used to directly detect CWs. Instead, individual pulsars can only be used to place upper limits on the strain amplitude of CWs. Currently, several pulsars have been observed with high cadence, including PSRs J1909\(-\)3744, J0437\(-\)4715, and J1713+0747, etc. These pulsars exhibit high timing precision and relatively short observation intervals, making them particularly well-suited for probing gravitational waves in the microhertz GW regime. 

The high-frequency limit of the sensitive band for PTA-based GW searches is widely assumed to the Nyquist rate \citep{Bracewell2000} associated with the average cadence of timing observations for individual pulsars in a PTA. Therefore, attempts at extending the high-frequency limit for resolvable GW sources are all based on high cadence observations of single pulsars. It should be noted that the actual cadences for pulsars in current PTAs vary over a large range and 1/(2 weeks) is more representative of its higher end.

Recent studies show that the high-frequency reach of PTAs for resolvable sources is not limited by the Nyquist rate of single-pulsar observations and the limiting frequency can be much higher than assumed so far. Higher frequency signal is preserved due to aliasing in the sequence of timing observations from each pulsar and can be unscrambled using asynchronous observations \citep{Bretthorst2001,Wong2006} from multiple pulsars. The lack of synchronicity, an inherent feature of PTA data, can be turned into an observational strategy, which we call staggered sampling, to boost the high-frequency reach of PTAs and bridge the gap in coverage of the GW spectrum between PTAs and LISA.

Exploring gravitational waves in the microhertz regime holds significant scientific value. This frequency band lies between the sensitive ranges of ground-based detectors (such as LIGO/Virgo/KAGRA) and traditional PTA observations, and is currently only accessible through high-cadence timing of MSPs. Detecting or constraining CWs in this regime would provide critical insights into the population of SMBHBs with sub-parsec separations, particularly those in the late inspiral phase prior to coalescence.

In the context of our results, a canonical system with $\mathcal{M}=10^{9}\,M_\odot$ radiating at $f=1$ \textmu Hz would generate a strain of $h_0\sim2\times10^{-14}$ at a luminosity distance of 100 Mpc. Our sky-averaged upper limit of $2.3\times10^{-13}$ therefore constrains only nearby ($\lesssim 9$ Mpc) binaries of this mass, while the optimal-sky sensitivity of $9\times10^{-14}$ extends this reach to $\sim22$ Mpc. Equivalently, at 100 Mpc our limits imply sensitivity to binaries with chirp masses $\gtrsim 2.5\times10^{9} M_\odot$. These quantitative bounds illustrate that our microhertz search directly probes the most massive and nearest SMBHBs that may exist in the late-inspiral regime.

The potential impact of solar-wind-induced dispersion-measure (DM) variations on our analysis has been carefully assessed. Solar-wind effects are expected to be most significant for pulsars located close to the ecliptic plane. PSR~J1909$-$3744 has a moderate ecliptic latitude and is therefore less strongly affected than low-latitude pulsars.

In our timing analysis, the solar-wind contribution was modeled using the default implementation in \texttt{tempo2}, with a fixed electron density parameter of $n_{\rm e}=4$. This value was not explicitly fitted but adopted as the standard default commonly used in timing analyses, and is now stated in Table \ref{tab:timing}.

Previous studies have shown that if the true solar-wind electron density differs from the assumed value, imperfect modeling can leave residual timing signatures. In particular, simulations presented in \citet{Liu_2025} demonstrate that even for relatively large deviations (e.g., a true solar-wind density corresponding to $n_{\rm e}=8$ while adopting $n_{\rm e}=6$ in the timing model), the resulting unmodeled timing residuals are typically at the level of $\sim100$~ns at L-band frequencies for pulsars with moderate ecliptic latitudes.

Our dataset further benefits from extensive multi-frequency coverage spanning approximately 800--3100~MHz, including simultaneous wideband observations between 800 and 1800~MHz. Such observations provide strong leverage to constrain chromatic DM variations. In addition, the deterministic timing model includes DM derivatives, and a stochastic DM noise component is incorporated in the noise model, further absorbing residual chromatic fluctuations.

Taken together, these considerations indicate that any residual solar-wind-induced timing perturbations arising from imperfect modeling are expected to be at or below the $\sim100$~ns level. Such residuals are well below the sensitivity relevant for the microhertz continuous gravitational-wave upper-limit analysis presented in this work and therefore do not significantly affect our results.



Furthermore, microhertz gravitational waves may also carry signatures of exotic astrophysical processes or alternative cosmological sources, such as cosmic strings. As such, searches for GW in this frequency range represent a unique opportunity to bridge the observational gap between nanohertz and millihertz GW detectors, and to open a new window into gravitational-wave astronomy.

\begin{acknowledgments}

This research was funded by the following grants: Major Science and Technology Program of Xinjiang Uygur Autonomous Region, grant number 2022A03013-4 to J.B.W.; Natural Science Foundation of Xinjiang Uygur Autonomous Region, grant number 2022D01D85 to J.B.W.; Tianshan Talents Program, grant number 2023TSYCTD0013 to J.P.Y., the National Key Research and Development Program of China (No.2022YFC2205203). We give thanks to Dr. Paul Baker for his helpful comments and suggestions.
\end{acknowledgments}

\bibliography{sample7}{}
\bibliographystyle{aasjournalv7}



\end{document}